# Superconductivity in $CoSr_2(Y_{1-x}Ca_x)Cu_2O_{7+\delta}$

Y. Morita, H. Yamauchi, M. Karppinen*

*Materials and Structures Laboratory, Tokyo Institute of Technology, Yokohama 226-8503, Japan*

**Abstract**

The roles of aliovalent $Ca^{II}$-for-$Y^{III}$ substitution and high-pressure-oxygen annealing in the process of "superconducterizing" the Co-based layered copper oxide, $CoSr_2(Y_{1-x}Ca_x)Cu_2O_{7+\delta}$ (Co-1212), were investigated. The as-air-synthesized samples up to $x = 0.4$ were found essentially oxygen stoichiometric ($-0.03 \le \delta \le 0.00$). These samples, however, were not superconductive, suggesting that the holes created by the divalent-for-trivalent cation substitution are trapped on Co in the charge reservoir. Ultra-high-pressure heat treatment carried out at 5 GPa and 500 °C for 30 min in the presence of $Ag_2O_2$ as an excess oxygen source induced bulk superconductivity in these samples. The highest $T_c$ was obtained for the high-oxygen-pressure treated $x = 0.3$ sample at ~40 K.

*Corresponding author: karppinen@msl.titech.ac.jp

The still expanding variety of high-$T_c$ superconductive and related $M_mA_2Q_{n-1}Cu_nO_{m+2+2n\pm\delta}$ ($M$-$m2(n-1)n$; category-A) and $M_mA_{2k}B_sCu_{1+k}O_{m+4k+2s\pm\delta}$ ($M$-$m(2k)s(1+k)$; category-B) phases are all based on rather robust $CuO_2$ planes stacked with other layer blocks that are more flexible in the cation and/or oxygen composition [1]. The *A* site is typically occupied by Ba, Sr or La, whereas the *Q* site prefers somewhat smaller cations of Ca, and/or Y and other (smaller) rare earths, *RE* (only for $n = 2$). In terms of the *M* constituent, the "chameleon" is nothing but the *M*-1212 structure: it tolerates a wide range of single-element constituents such as Cu, Co, Fe, Nb, Ta, Ru, Hg, Tl, Al, Ga, C, and various cation mixtures as *M*. Some of the *M*-1212 phases are well-established superconductors, whereas others remain to be "superconducterized" yet.

The amount and location of oxygen in the $M\mathrm{O}_{1\pm\delta}$ charge reservoir of the $M$-1212-type phases varies considerably. In the prototype $M$-1212 phase, *i.e.* $CuBa_2YCu_2O_{7-\delta}$ (Cu-1212), the excess oxygen atoms fill half of the perovskite-type positions such that the charge-reservoir Cu atom with surrounding oxygen atoms forms a square-planar $CuO_4$ polyhedron. Octahedral coordination is possible as well, with excess oxygen atoms being located at either rock-salt position for, *e.g.* $M$ = Tl [2], or perovskite position for, *e.g.* $M$ = Ru [3], though the real structures deviate from the ideal ones: in Tl-1212 the oxygen site in the TlO layer is subject to large concentration of oxygen vacancies, while in Ru-1212 the $RuO_6$ octahedra are rotated such that the oxygen atom is slightly shifted from the ideal $a$-axis site. Another well-known example of the $M$-1212 phase with a rock-salt type $M\mathrm{O}_{1\pm\delta}$ layer, *i.e.* Hg-1212 [4], exhibits even much larger oxygen deficiency than Tl-1212, such that the coordination polyhedron is considered as an O-Hg-O dumbbell. The tetrahedral coordination polyhedron has been seen for three phases with $M$ = Ga [5], Al [6,7] and Co [8,9]. Among these, Ga-1212 shows superconductivity up to ~70 K when properly doped with holes [10,11]. Superconductivity in Al-1212 has been somewhat debated [12,13]; even in the positive case the onset temperature is low, *i.e.* 18 ~ 30 K [13]. The third phase, Co-1212, was synthesized in 1989 with the composition of $CoSr_2YCu_2O_{7+\delta}$ [14]. The crystal structure of $CoSr_2YCu_2O_{7+\delta}$ was later reported by Huang *et al.* [8]. The charge reservoir of Co-1212 consists of zigzag chains of corner-linked $CoO_4$ tetrahedra that run diagonally relative to the perovskite base. Later this picture was further elaborated by careful ED and TEM studies that revealed the presence of an orthorhombic superstructure originating from a regular alternation of two zigzag chains that are mirror images of each other [15,16]. Even though the Co-1212 phase has been well known for long, the first indication of superconductivity at ~30 K was revealed for it only very recently by Mochiku *et al.* [17]. In the present communication we summarize our systematic work carried out independently of that of Mochiku *et al.* on the effects of aliovalent $Ca^{II}$-for-$Y^{III}$ substitution and high-oxygen-pressure annealing in $CoSr_2YCu_2O_{7+\delta}$. The work resulted in finding of bulk superconductivity with $T_c$ up to 40 K.

We first synthesized a series of samples of $CoSr_2(Y_{1-x}Ca_x)Cu_2O_{7+\delta}$ ($x$ = 0 ~ 0.7) *via* solid-state reaction in air. The starting materials, $Co_3O_4$, $SrCO_3$, $Y_2O_3$, $CaCO_3$ and CuO, were mixed to nominal ratios and calcined at 975 $^{o}$C and 1000 $^{o}$C with an intermediate grinding. The phase purity was checked for each sample with powder x-ray diffraction (XRD; MAC Science: MXP18VAHF[22]; Cu $K_\alpha$ radiation). Figure 1 displays XRD patterns for the as-synthesized samples (henceforth referred to as AS samples) with various Ca contents, $x$. From Fig. 1, the AS samples are free of any visible impurity peaks at least up to $x = 0.5$. For the samples with $x \geq 0.5$, a small unidentified diffraction peak develops around $2\theta \approx 44°$ and also the peak at $2\theta \approx 32°$ increases abruptly. The lattice parameters as determined from the XRD data (using space group *Imcm* [8]) are plotted against $x$ in Fig. 2. (Note that we use the crystal-axis naming scheme typically employed for the structures of high-$T_c$ superconductors, *i.e.* assign the longest lattice parameter along the layer-piling direction as $c$ axis.) From Fig. 2, systematic decrease in the $a$- and the $b$-axis parameter and increase in the $c$-axis parameter are evident with increasing Ca-substitution level up to $x = 0.5$. Beyond $x = 0.5$, the $c$ parameter still continues to increase but the behavior of parameters, $a$ and $b$, is different from that seen for $x \leq 0.5$. Therefore it is likely that the solubility limit of Ca in $CoSr_2(Y_{1-x}Ca_x)Cu_2O_{7+\delta}$ is in the vicinity of $x = 0.5$. The fact that the $c$-axis parameter increases with increasing Ca-substitution level is a clear indication of Ca entering the Y site, since substitution of Sr by the smaller alkaline earth, Ca, would result in shrinkage of the $c$ axis. Related experimental data to support this reasoning are reported for the $AlSr_2HoCu_2O_7$ phase [18]. In $AlSr_2HoCu_2O_7$, Ca was found to enter either the Sr site or the Ho site depending on the synthesis condition: substitution of Sr ended up with a shortened $c$ axis, whereas that of Ho resulted in a lengthened $c$ axis.

When aliovalent cation substitution such as the present $Ca^{II}$-for-$Y^{III}$ one is applied, it is highly probable that the lattice responses to the “substitutional oxidation” by concomitantly reducing the oxygen content [19]. Therefore, we precisely analyzed the oxygen content of each AS sample. Iodometric titration has been shown to be a well-suited method for perovskite Co oxides [20,21], and was therefore employed for the present oxygen-content analyses. The titration was carried out in an air-tight cell under flowing $N_2$ gas. The sample (of *ca.* 30 mg) was dissolved in oxygen-freed 1 M HCl solution containing an excess of KI. Reaction with iodide results in reduction of all high-valent Cu and Co species of the sample, *i.e.* $Cu^{II/III}$ and $Co^{III/IV}$, to monovalent copper (solid CuI) and divalent cobalt ($Co^{2+}$ ions), respectively. Concomitantly stoichiometric amount of iodine is formed, that is then titrated with $Na_2S_2O_3$ solution (*ca.* 0.015 M), using starch as an indicator for visual end-point detection. The values of oxygen content determined for the AS samples with various Ca contents are given in Table 1. From Table 1, it is seen that the oxygen content remains essentially stoichiometric ($-0.03 \leq \delta \leq 0.00$) up to $x = 0.4$. We have thus confirmed that up to $x = 0.4$, *i.e.* as long as the oxygen content remains constant, the $Ca^{II}$-for-$Y^{III}$ substitution increases the oxidation state of Cu and/or Co, thus working in the way expected for ideal aliovalent cation substitution.

For a non-superconductive Ca-free $CoSr_2YCu_2O_{6.97}$ sample our preliminary Cu $L$-edge XANES (x-ray absorption near-edge structure) spectroscopy study showed that the valence of Cu in this sample is II [22]. Accordingly, based on a simple charge-balance calculation, the valence of Co is III. Now, taking respectively the values of II and III as the “bottom valence values” for Cu and Co in an undoped situation, we may calculate the amount of excess positive charge (holes) *per* formula unit in the present $CoSr_2(Y_{1-x}Ca_x)Cu_2O_{7+\delta}$ samples from the nominal cation content and the amount of oxygen precisely analyzed by iodometric titration. The hole numbers thus derived are given in Table 1. The amount of induced holes increases with increasing $x$ up to $x = 0.4$, where it reaches a value of 0.36 holes *per* formula unit.

Next we checked possible superconductivity for the AS samples. Magnetization measurements were performed with a SQUID magnetometer (Quantum Design: MPMS-XL) in both field-cooled (FC) and zero-field-cooled (ZFC) modes down to 5 K. None of the AS samples showed any sign of superconductivity, being in agreement with the observation made earlier for an $x = 0.4$ sample by Huang *et al.* [8]. In Fig. 3(a) displayed is the $\chi$ *vs.* $T$ curve for the $x = 0.4$ sample as an example. As given in Table 1, the amount of excess holes varies from 0.14 to 0.36 *per* formula unit in the $x = 0.2$ - 0.4 samples. Assignment of all the holes to the $CuO_2$ plane (that appears in duplicate in the formula unit) would end up with the $CuO_2$-plane hole concentration ranging from 0.07 to 0.18. These values should be adequate for inducing high-$T_c$ superconductivity. Here, to explain the absence of superconductivity we recognize the following two plausible reasons: (*i*) Co and Cu are partly intermixed such that the content of Co in the $CuO_2$ plane is high enough to kill superconductivity [8], and (*ii*) holes induced by Ca substitution are not directed into the $CuO_2$ plane but are (mostly) located in the $CoO_{1+\delta}$ charge reservoir. Meanwhile we assume that the former is not the present case but the latter applies, that is, the induced holes are trapped on $Co^{III/IV}$ in the $CoO_{1+\delta}$ charge reservoir.

In order to push the holes into the $CuO_2$ plane and/or introduce more holes in the lattice we performed a high-pressure-oxygen (HPO) annealing for the AS samples. The HPO annealing was carried out in a cubic-anvil-type ultra-high-pressure apparatus at 5 GPa and 500 $^{o}$C for 30 min. As an oxygen source, $Ag_2O_2$ (that readily releases $O_2$ upon heating while decomposing *via* $Ag_2O$ to Ag) was utilized: $Ag_2O_2$ was mixed with Co-1212 powder to a molar ratio of 1 : 2. XRD patterns for the HPO samples are shown in Fig. 4. Comparison with those for the AS samples (Fig. 1) reveals that the HPO annealing had no effect on the phase purity of the samples. On the other hand, in Fig. 2 where the lattice parameters are plotted for both the AS and the HPO samples, it is observed that upon HPO annealing the lattice drastically shrank along the $b$ and $c$ axes, whereas the $a$-axis parameter remained unchanged. As a consequence of the $b$-axis shrinkage, the degree of orthorhombic distortion decreases. We recognize the same from the fact that every pair of double peaks seen in the XRD pattern for the AS samples, *e.g.* 026 and 206, 136 and 316, 138 and 318, merges into a single peak after HPO annealings. This suggests that the $CoO_4$ tetrahedra have got broken as a consequence of the change in the oxygen amount/arrangement about the Co atom. For the HPO-annealed samples superconductivity appears with the transition temperature ($T_c$) up to 40 K (as determined at the onset of the diamagnetic signal) for the $x = 0.3$ sample. The largest volume fraction of superconductivity (16 % from the ZFC data and 11 % from the FC data) is due to the $x = 0.35$ sample. In Fig. 5, we show the $T_c$ values for all the HPO samples as plotted against $x$. A bell-shape relation is seen. In Fig. 3(b) we have shown the $\chi$ *vs.* $T$ curve for the $x = 0.4$ sample as an example.

Now a question arises on the primary reason for the appearance of superconductivity in the HPO samples. To be able to reach the correct answer, we first investigated whether the pure high-pressure (HP) effect alone would induce superconductivity in the Ca-substituted $CoSr_2(Y_{1-x}Ca_x)Cu_2O_{7+\delta}$ samples. As-synthesized powder with $x = 0.4$ was annealed without $Ag_2O_2$ at 5 GPa and 500 °C for 30 min, *i.e.* under the condition otherwise the same as that employed for the HPO samples except that no excess oxygen source was added. Superconductivity did not show for this HP sample. Furthermore, upon the HP annealing the lattice parameters of the AS sample as well as the $\chi$ *vs.* $T$ behavior (see Fig. 3(c) for the $x = 0.4$ case) remained unchanged. It is thus concluded that both the drastic decrease in the lattice parameters, $b$ and $c$, and the appearance of superconductivity upon high-pressure annealing have resulted from the application of high oxygen partial pressure during annealing.

Next we annealed the HPO-superconducterized $x = 0.4$ sample in flowing $N_2$ gas. This annealing was carried out in a thermobalance (Perkin Elmer: Pyris 1) in order to be able to thermogravimetrically (TG) follow changes in the sample weight occurring during the annealing. The initial mass of the sample was 22 mg and the heating rate was 1 °C/min. After reaching the final temperature of 500 °C the sample was quenched to room temperature. From the TG curve recorded, two weight-loss steps are seen about 150 and 300 °C (Fig. 6). The overall amounts of oxygen depleted in these two steps from the HPO sample (that is a mixture of $CoSr_2(Y_{0.6}Ca_{0.4})Cu_2O_{7+\delta}$ + $(1-y)$ Ag + $0.5y$ $Ag_2O$) were calculated at ~0.4 and ~0.2, respectively. From the XRD pattern for the HPO-$N_2$ sample it was concluded that the $N_2$ annealing made the lattice parameters of the Co-1212 phase to recover back to the same values as those of the AS sample before the HPO annealing. On the other hand, the amount of Ag seen in the XRD pattern did not visibly increase as compared to that seen for the HPO sample. Therefore, it is likely that the weight/oxygen loss detected by TG upon $N_2$ annealing is primarily due to the Co-1212 phase itself. This conclusion is supported by the magnetization data: upon annealing the HPO sample in $N_2$ not only superconductivity but also the magnetic transition characteristically seen for all the HPO samples around 150 K disappeared (Fig. 3(d)). Note that the origin of the magnetic transition (manifested by the branching of the ZFC and FC magnetization curves) was left unaddressed yet, as the structure of the $CoO_{1+\delta}$ charge reservoir is not determined for the HPO samples. The second, smaller weight step in the TG curve about 300 °C well coincides with that seen for the decomposition of pure $Ag_2O$ [23]. We thus assign the second step for the decomposition of leftover $Ag_2O$, *i.e.* $y \approx 0.4$, and the first step only to the oxygen evolution from the Co-1212 phase, *i.e.* $\delta \approx 0.4$.

As the final step of the present work we checked the optimum amount of $Ag_2O_2$ used as a source of excess oxygen in the HPO annealing. The first HPO treatments were all carried out with 50 mol-% $Ag_2O_2$. Here we made *ad hoc* HPO annealings for the $x = 0.4$ sample using 25 and 100 mol-% $Ag_2O_2$. It was found that with 25 mol-% $Ag_2O_2$, the $c$ axis decreased less as compared with the case of 50 mol-%. More importantly, the sample did not show superconductivity. On the other hand, with 100 mol-% $Ag_2O_2$ peaks due to $Ag_2O$ appeared in the XRD pattern for the HPO annealed sample. The annealed sample exhibited superconductivity with a lower $T_c$ value of 20 K than that of the HPO annealed product with 50 mol-% of $Ag_2O_2$ ($T_c \approx 35$ K). It was thus revealed that the optimum amount of $Ag_2O_2$ as an oxygen source for the HPO annealing is in the vicinity of 50 mol-%. Here it should also be noted that we annealed the two new HPO samples in a thermobalance in $N_2$. From the TG curves shown in Fig. 6 together with that for the sample with 50 % $Ag_2O_2$ we could see the following features: the second weight-loss step about 300 $^{o}$C that we had previously assigned to the decomposition of leftover $Ag_2O$ was not seen at all for the HPO sample with 25 % $Ag_2O_2$ ($y \approx 0$), whereas it was more pronounced for that with 100 % $Ag_2O_2$ ($y \approx 1.0$) in comparison with the original HPO sample with 50 % $Ag_2O_2$ ($y \approx 0.4$). The amount of oxygen depleted from the Co-1212 phase in the first weight-loss step increased with increasing amount of $Ag_2O_2$ used for the HPO annealing, *i.e.* $\delta \approx 0.3$ (25 % $Ag_2O_2$), 0.4 (50 %) and 0.5 (100 %).

In conclusion, we have systematically clarified the influence of aliovalent $Ca^{II}$-for-$Y^{III}$ substitution and high-pressure-oxygen annealing on the superconductivity of $CoSr_2(Y_{1-x}Ca_x)Cu_2O_{7+\delta}$ (Co-1212) samples. It was found that at least up to $x \approx 0.4$ Ca dissolves in the Co-1212 phase at the Y site. Within this range of Ca substitution the oxygen content of as-synthesized samples remained stoichiometric. Therefore we could assure that holes were gradually doped in the Co-1212 structure upon Ca substitution. The as-synthesized samples, however, were not superconductive, suggesting that the holes created by the divalent-for-trivalent substitution are trapped about the Co atom in the charge reservoir. The combination of Ca substitution and high-pressure-oxygen annealing was necessary to induce superconductivity in the Co-1212 phase. The highest $T_c$ obtained in the present study for the HPO-annealed $x = 0.3$ sample was ~40 K.

We gratefully acknowledge Dr. V.P.S. Awana for his contribution in initiating the research on the phases of Co-12*s*2 homologous series in our group. Dr. T. Motohashi is acknowledged for fruitful discussions. This work was supported by Grants-in-aid for Scientific Research (Nos. 15206002 and 15206071) from Japan Society for the Promotion of Science.

**Table 1**. The degree of oxygen nonstoichiometry, δ, as determined by iodometric titration and the amount of "excess holes" *per* formula unit for the as-synthesized (AS) $CoSr_2(Y_{1-x}Ca_x)Cu_2O_{7+\delta}$ samples. The amount of "excess holes" is calculated from the values of $x$ and δ against the "bottom valence values" of II and III for Cu and Co, respectively.

| ***x*** | **δ** | **"excess holes"** |
|---|---|---|
| 0.0 | -0.03 | -0.06 |
| 0.1 | -0.03 | 0.04 |
| 0.2 | -0.03 | 0.14 |
| 0.3 | 0.00 | 0.30 |
| 0.4 | -0.02 | 0.36 |
| 0.5 | -0.10 | 0.30 |
| 0.6* | -0.20 | 0.20 |
| 0.7* | -0.27 | 0.16 |

*May not be totally phase-pure.

## Figure captions

**Fig. 1.** XRD patterns for the as-synthesized (AS) samples of $CoSr_2(Y_{1-x}Ca_x)Cu_2O_{7+\delta}$. Note that the intensity scale is logarithmic.

**Fig. 2.** Lattice parameters, *a*, *b* and *c*, for the as-synthesized (AS; open circles) and the high-pressure-oxygen annealed (HPO; filled triangles) samples of $CoSr_2(Y_{1-x}Ca_x)Cu_2O_{7+\delta}$.

**Fig. 3.** Magnetic susceptibility ($\chi$) *versus* temperature ($T$) curves measured in FC and ZFC modes for the following $x = 0.4$ samples of $CoSr_2(Y_{1-x}Ca_x)Cu_2O_{7+\delta}$: as-synthesized (AS), high-pressure-oxygen (HPO) annealed, high-pressure (HP) annealed, and $N_2$ annealed after HPO annealing (HPO-$N_2$).

**Fig. 4.** XRD patterns for the high- pressure-oxygen (HPO) annealed samples of $CoSr_2(Y_{1-x}Ca_x)Cu_2O_{7+\delta}$. Besides those for the Co-1212 phase, seen are peaks due to Ag as a decomposition product of $Ag_2O_2$. Note that the intensity scale is logarithmic.

**Fig. 5.** Superconductivity transition temperature, $T_c$, for the high- pressure-oxygen (HPO) annealed samples of $CoSr_2(Y_{1-x}Ca_x)Cu_2O_{7+\delta}$.

**Fig. 6.** TG curves for the $N_2$ annealing of the high-pressure-oxygen (HPO) annealed $x = 0.4$ samples of $CoSr_2(Y_{1-x}Ca_x)Cu_2O_{7+\delta}$: for the HPO annealing 25 %, 50 % or 100 % $Ag_2O_2$ was used. The heating rate is 1 $^{o}$C/min. The first/second weight-loss step is due to oxygen evolution from the Co-1212 phase/decomposition of leftover $Ag_2O$.

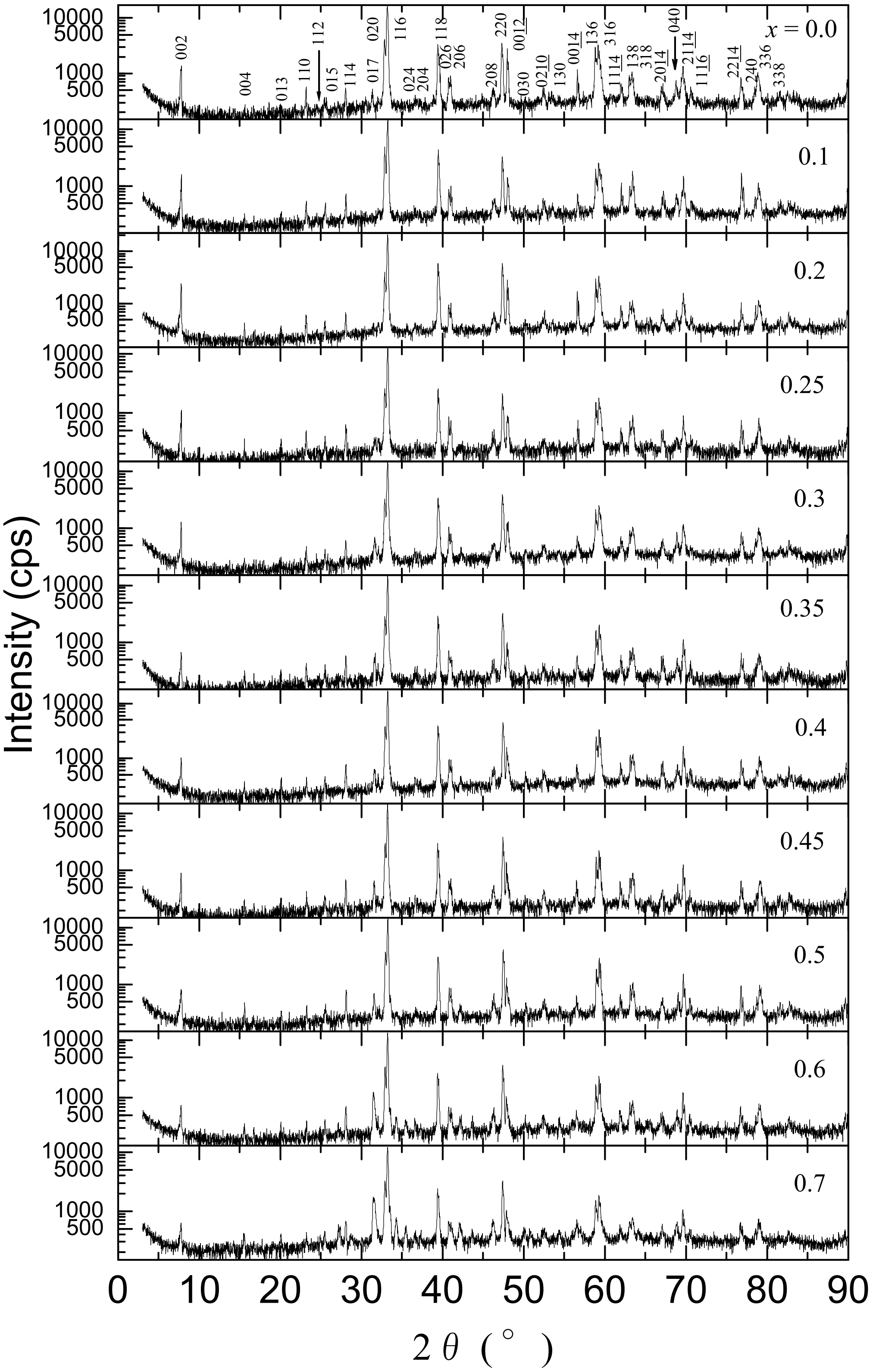

Morita et al: Fig. 1.

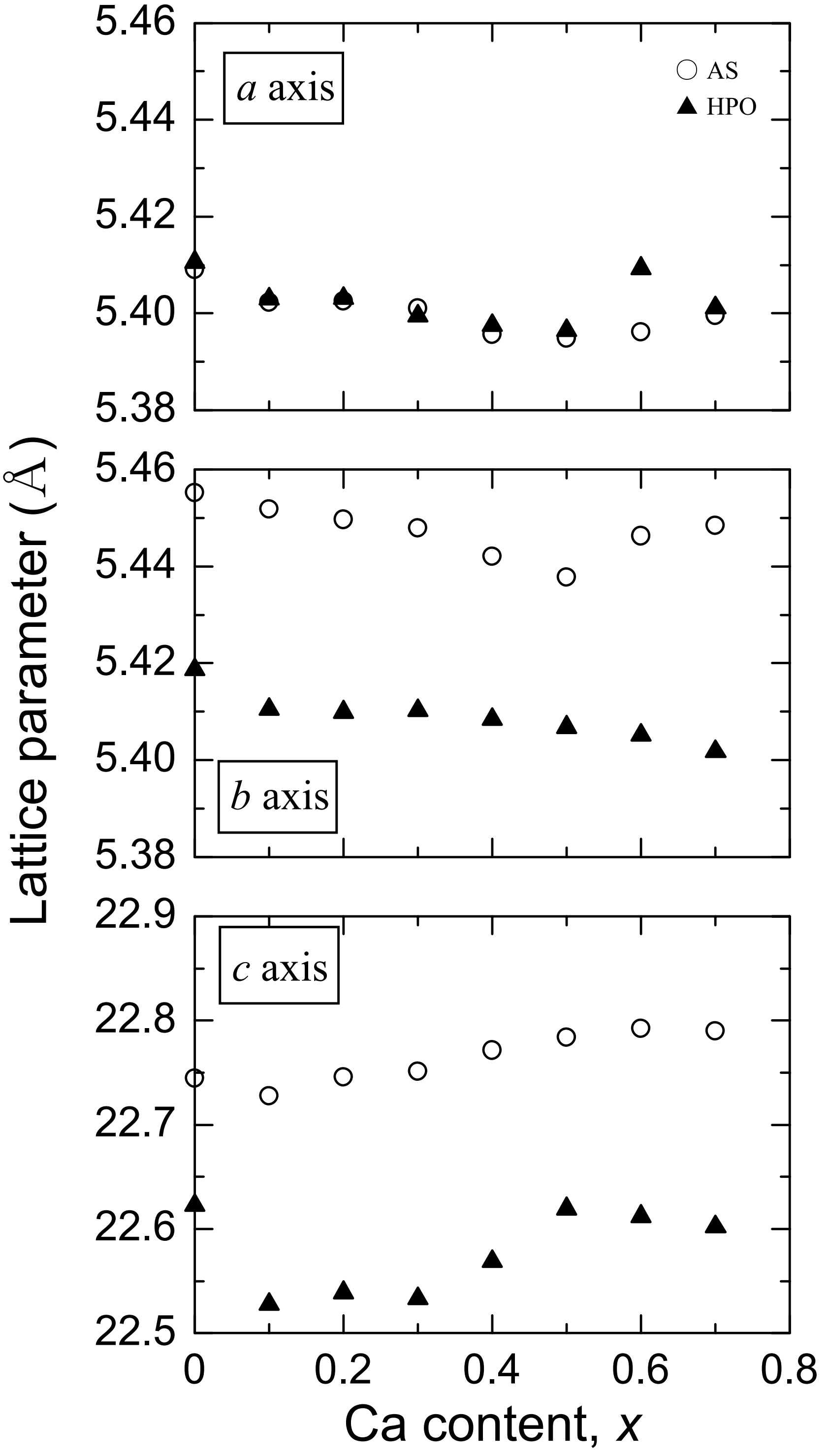

Morita et al: Fig. 2.

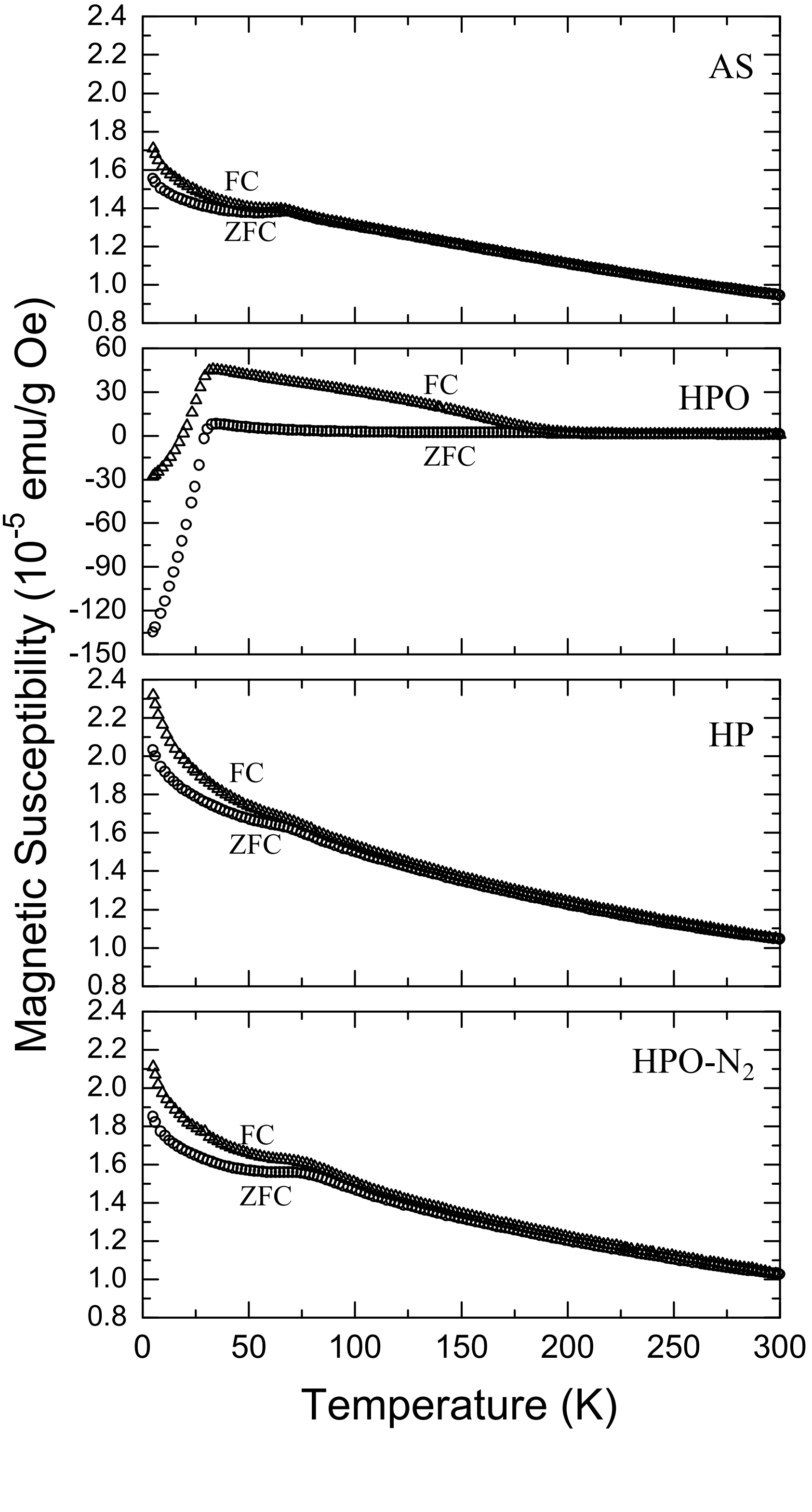

Morita et al: Fig. 3.

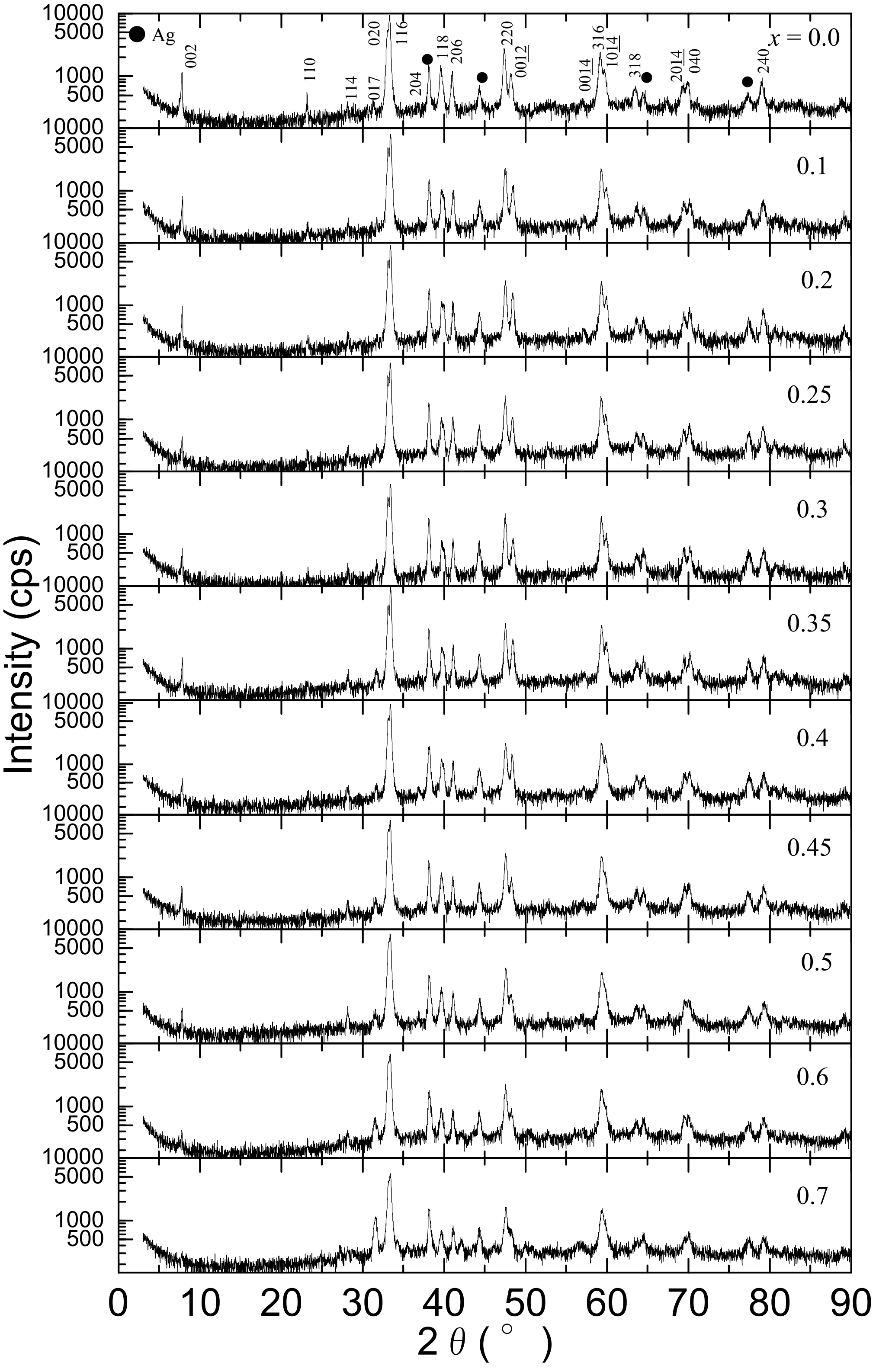

Morita et al: Fig. 4.

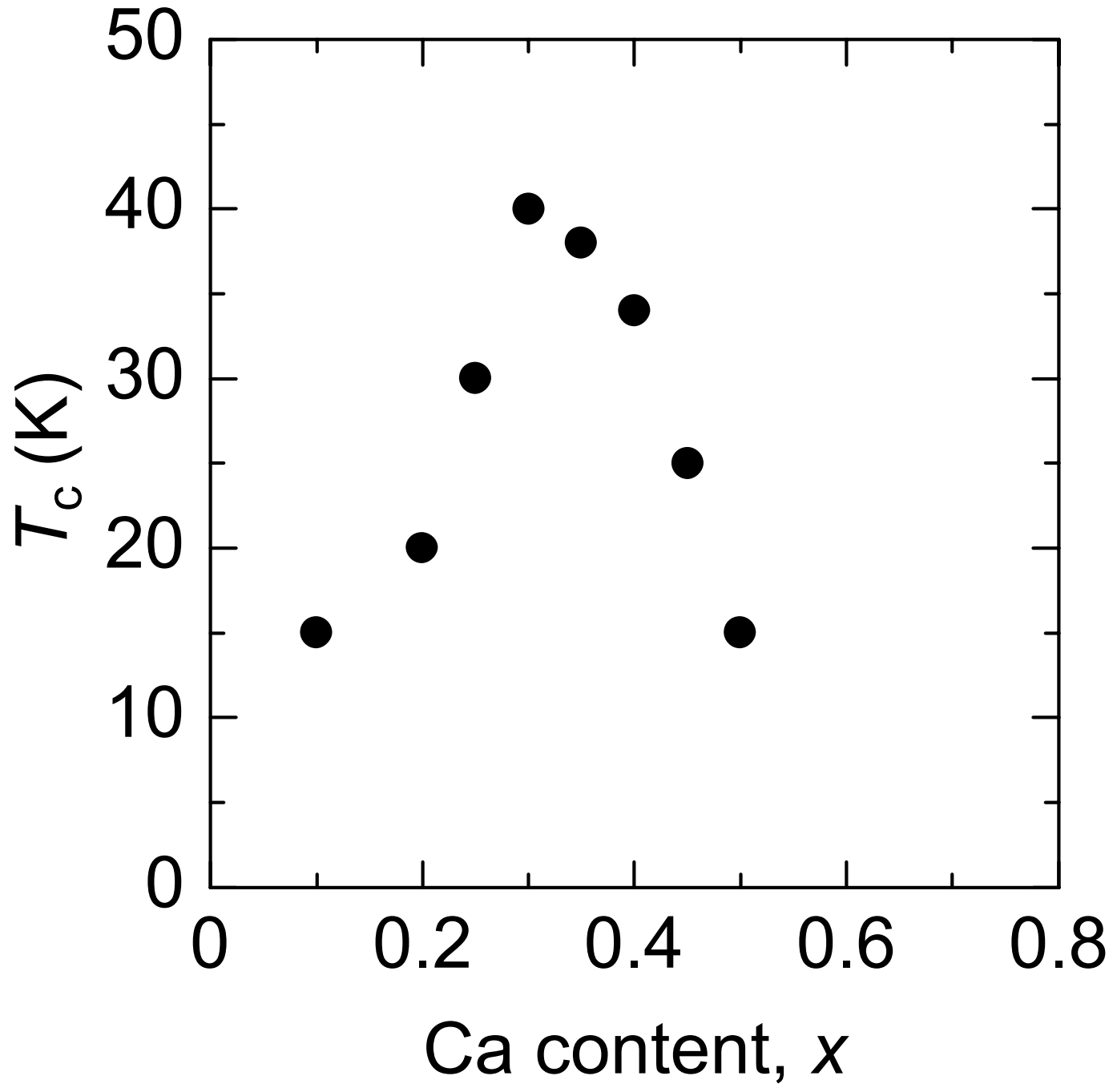

Morita et al: Fig. 5.

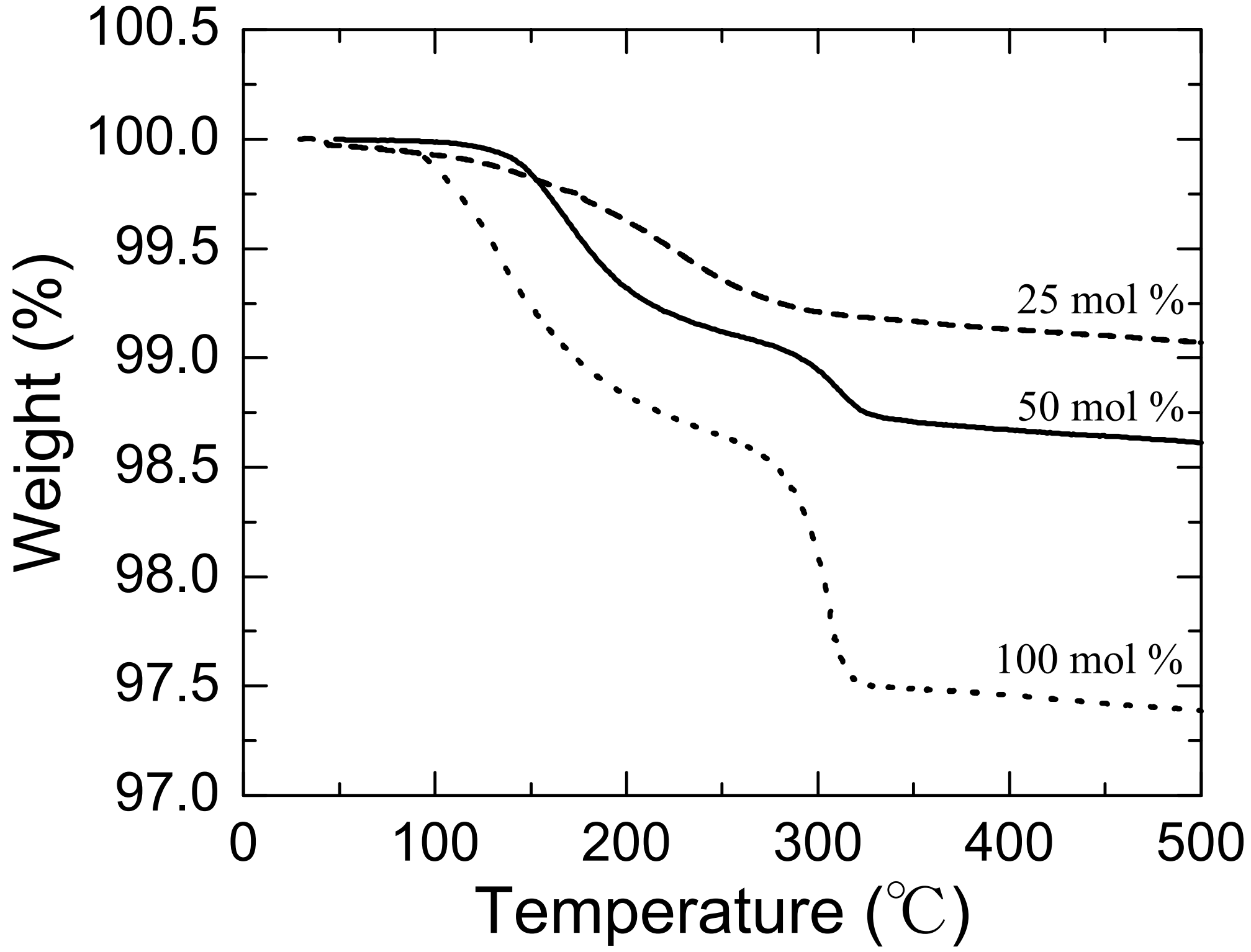

Morita et al: Fig. 6.